\documentclass[twocolumn,showpacs,preprintnumbers,amsmath,amssymb,prl,floatfix]{revtex4}
\usepackage{graphicx}
\usepackage{amssymb,amsfonts,amsmath}
\usepackage{xcolor}

\def\kT{{k_{B}T}}

\begin{document}

\title{Mobility Measurements Probe Conformational Changes\\ in Membrane 
Proteins due to Tension}

\author{Richard G.~Morris and Matthew S.~Turner}

\affiliation{Department of Physics and Centre for Complexity Science, 
University of Warwick, Coventry, CV4 7AL, UK.}

\begin{abstract}
	The function of membrane-embedded proteins such as ion channels depends 
	crucially on their conformation.  We demonstrate how conformational changes 
	in asymmetric membrane proteins may be inferred from measurements of their 
	diffusion.  Such proteins cause local deformations in the membrane, which 
	induce an extra hydrodynamic drag on the protein.  Using membrane tension 
	to control the magnitude of the deformations and hence the drag, 
	measurements of diffusivity can be used to infer--- via an elastic model of 
	the protein--- how conformation is changed by tension.  Motivated by recent 
	experimental results [Quemeneur {\it et al.}, Proc.~Natl.~Acad.~Sci.~USA, 
	{\bf 111} 5083 (2014)] we focus on KvAP, a ubiquitous voltage-gated 
	potassium channel.  The conformation of KvAP is found to change 
	considerably due to tension, with its `walls', where the protein meets the 
	membrane, undergoing significant angular strains.  The torsional stiffness 
	is determined to be $26.8\,\kT$ at room temperature.  This has implications 
	for both the structure and function of such proteins in the environment of 
	a tension-bearing membrane.
\end{abstract}


\maketitle

Recently, Quemeneur {\it et al.}~\cite{FQ+13} measured how the diffusion of 
KvAP was affected by membrane tension.  KvAP is an example of a protein that is 
found to have an affinity for curved membranes~\cite{SA+13}, implying an 
asymmetric, cone-like shape. The protein induces a localised deformation, or 
`dimple', in the membrane, the magnitude (and extent) of which decreases as the 
applied tension is increased.  To investigate the effect of shape on dynamics, 
the authors of \cite{FQ+13} traced the motion of KvAP at different membrane 
tensions and compared the corresponding diffusion constant to the reference, or 
control, values exhibited by a cylindrically shaped protein (of equivalent 
radius), which can be related to the theory of Saffman and Delbr\"{u}ck 
\cite{PGSMD75}.  At high tensions the corrections due to the shape of KvAP were 
very small ($\sim$ 5\%), whilst at lower tensions the corrections ($\sim$ 40\%) 
were much more pronounced.

In order to explain these results, the authors of \cite{FQ+13} invoked a 
polaron-like theory \cite{AN+09,ER-G+10,VDDSD10}.  This involves adding an 
extra term to the Hamiltonian of the membrane, which is coupled locally to 
membrane curvature and gives rise to a dimple consistent with the protein's 
shape.  An Oseen approximation is then used to calculate an additional drag, 
which arises because a moving dimple must displace the surrounding viscous 
fluid.  The corresponding reduction to the diffusion constant is then found by 
using the Stokes-Einstein relation.  However, the approach neglects 
(\textit{i}) the fact that membranes are themselves incompressible fluids, 
satisfying a two-dimensional form of Stokes equation, and (\textit{ii}) that 
the movement of the protein imposes particular boundary conditions on the 
membrane flow (and the membrane flow, in turn, imposes conditions on the 
surrounding fluid flow).  Moreover, the additional drag calculated in 
\cite{FQ+13} was found to be too small to explain the experimental data, 
leading the authors to explore additional dissipative mechanisms.  These were 
traced to membrane shear flows, or to the assumption that a protein might drag 
a large island of immobilised lipids through the membrane.  However, the 
effects of these modifications were calculated within the same Oseen 
approximation, and cannot be expected to reliably describe any properties 
related to membrane flows for the reasons given: such flows must satisfy the 
equations of two-dimensional incompressible Stokes flow, and are subject to 
appropriate physical boundary conditions near the moving object.  It is for 
these reasons that the results of Saffman and Delbr\"{u}ck do not emerge in the 
appropriate limit of zero curvature in \cite{FQ+13}.

Here, we instead seek a classical hydrodynamic explanation for the additional 
drag, and hence reduced diffusion, of curvature-inducing proteins.  In order to 
take account of the geometry of the membrane, we employ a covariant formulation 
of low Reynolds number hydrodynamics in two 
dimensions~\cite{DRDandMST,MLH+07,MLHAJL10,MAAD09}.  In doing so, we neglect 
both membrane fluctuations and any chemical interactions occurring between the 
protein and the amphiphilic molecules that make up the membrane 
\cite{YGJ-LP03,DVTCMD12}.  By treating the membrane hydrodynamics in this way 
we find no additional dissipative mechanisms are required.

\begin{figure*}
\centering
\includegraphics[width=0.97\textwidth]{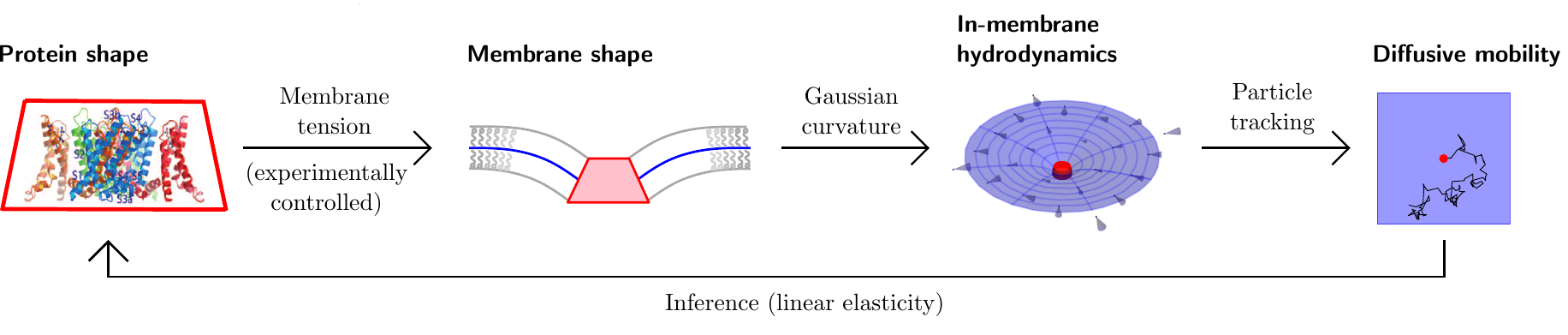}
\caption
{
	{\bf Flow diagram.}  The shape of KvAP induces a local deformation in the 
	membrane, resulting in non-zero Gaussian curvature in the vicinity of the 
	protein.  As tension is applied to the membrane, the deformation becomes 
	more localised, increasing Gaussian curvature.  A covariant formulation of 
	low Reynolds number hydrodynamics demonstrates that Gaussian curvature 
	increases the drag on the protein, therefore reducing diffusion.  As a 
	result, measurements of particle trajectories, such as those of 
	Ref.~\cite{FQ+13}, can be used alongside a simple elastic model of protein 
	deformation to infer how protein shape is changed by applied tension.
}
\label{fig:combined}
\end{figure*}

If the shape of the protein is fixed, our calculations predict an {\it 
increased} hydrodynamic drag at high tensions.  The reason is that the induced 
dimple in the membrane becomes localised and sharp, increasing the Gaussian 
curvature of the membrane in the vicinity of the protein and introducing 
additional hydrodynamic shear stresses (see, for example, Ref.~\cite{MAAD09}).  
Such an effect is not apparent in the data, which suggests that, for 
sufficiently high tensions, the Brownian motion of KvAP should be 
indistinguishable from a cylindrically shaped protein of the same radius (such 
as the aquaporin AQP0, used as a control by Quemeneur {\it et al.}).  This is 
evidence that the conformation, or shape, of the protein is changed by the 
torque exerted on the `walls' where it meets the membrane
\footnote{Notwithstanding membrane hydrodynamics, the polaron model of 
	\cite{FQ+13} also permits protein deformation.}.  Combining our 
	hydrodynamic theory with linear elastic response yields an excellent fit to 
	the data \cite{FQ+13} and predicts the relevant torsional stiffness of 
	KvAP.  A flowchart representing our approach is shown in 
	Fig.~\ref{fig:combined}.

\begin{figure}[b!]
\centering
\includegraphics[width=0.5\textwidth]{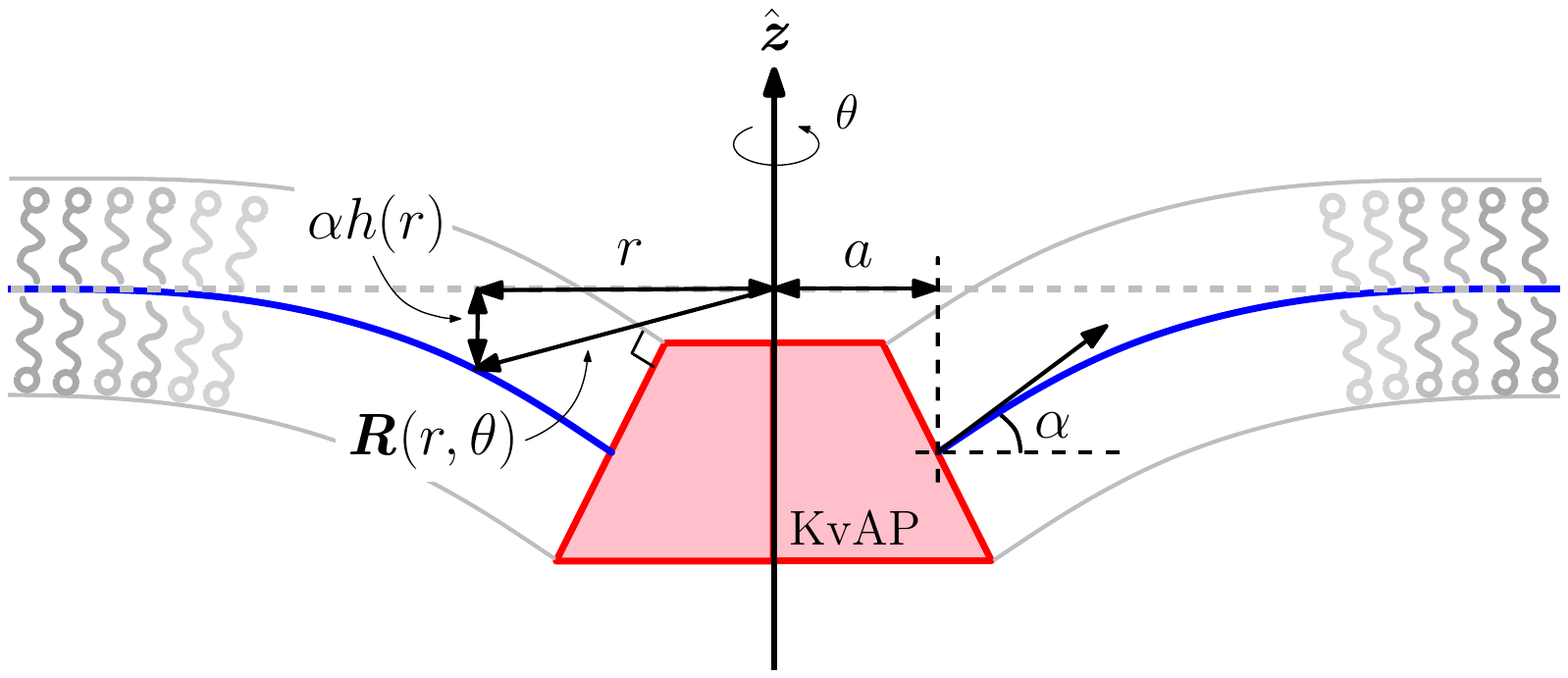}
\caption
{
	{\bf Sketch.} The embedded membrane protein KvAP induces a local curvature 
	in an otherwise planar membrane.  The mid-plane of the membrane is 
	characterised by a cylindrically symmetric height $h(r),\ \forall\ 
	r\in[a,\infty)$ and is further proportional the contact angle $\alpha$, 
		which also serves as the small parameter in our perturbation theory for 
		the hydrodynamic drag acting on KvAP.
}
\label{fig:schematic}
\end{figure}

To develop a theory for the hydrodynamics associated with the motion of KvAP, 
the induced shape of the membrane must first be calculated.  Taking the 
mid-plane of the bilayer to be a smooth Reimannian manifold $\mathcal{S}$, each 
point on $\mathcal{S}$ is attributed a Helfrich-like free energy per unit 
area~\cite{WH73,US96}.  The lipids are assumed to remain well ordered 
everywhere and therefore the bilayer has a bending energy of $2\kappa H^2$, 
where $\kappa$ is a constant and $H$ is the mean curvature.  The spontaneous 
curvature is zero, and the membrane is under lateral tension $\sigma$.  In the 
experiments of \cite{FQ+13}, this is controlled by the pressure difference 
between the interior and exterior of a giant unilamellar vesicle.  Neglecting 
fluctuations, the shape of the membrane at equilibrium is then found by 
minimising the total free energy
\begin{equation}
	E=\int_\mathcal{S} \left(2\kappa\,H^2 + \sigma\right)dA,
	\label{eq:E}
\end{equation}
where $dA$ is used as a shorthand for the volume 2-form, $\mathrm{vol}^2$, 
associated with $\mathcal{S}$.  Using a small angle approximation, the solution 
can be characterised by an axisymmetric height field $\alpha\,h(r),\ \forall\  
r\in [a,\infty)$, where $a$ is the radius of the protein and $\alpha$ is the 
	contact angle subtended at the walls of the protein (see 
	Fig.~\ref{fig:schematic}).  Up to a constant factor, the variational 
	procedure yields an order-$0$ modified Bessel function of the second kind 
	(see \cite{TRW+98} and SI):
\begin{equation}
	h(r) = l\, K_0 \left( r/l \right)\,/\,K_1 \left( a/l\right),
	\label{eq:h}
\end{equation}
where $l=\sqrt{{\kappa}/{\sigma}}$ is the membrane correlation length.  Notice 
that increasing the surface tension leads to an increasingly localised membrane 
deformation, or dimple (see Fig.~1 of the SI).

The effect of the induced-shape (\ref{eq:h}) on protein diffusion may be 
calculated by first computing the hydrodynamic drag, $\lambda$, on a protein 
moving with constant velocity, and then relating this to the diffusion constant 
via the fluctuation-dissipation theorem~\cite{Batchelor}.  We consider the 
protein moving laterally ({\it i.e.}, perpendicular to the $z$-axis of 
Fig.~\ref{fig:schematic}) with a velocity whose magnitude $V$ is sufficiently 
small that $h(r)$ remains good approximation to the membrane shape \footnote
{
	Lateral motion of the protein generates a spatially varying two-dimensional 
	pressure ({\it i.e.}, tension) in the membrane, as well as a 
	rotation-inducing torque on the protein.  However, the corrections to 
	$h(r)$, which result in an asymmetric shape, are proportional to $V$.  
	Since the drag coefficient $\lambda$ is the leading order coefficient in a 
	power series expansion of $F$ in terms of $V$, it must not contain any such 
	corrections, and the stationary membrane profile suffices for its 
	calculation.  This not the case in \cite{FQ+13}, which concerns the 
displacement of the surrounding fluid, and hence an asymmetric profile is 
necessary for force balance.} and the hydrodynamics remains at low Reynolds 
number \footnote
{
	Two dimensional fluids, such as membranes, are known to experience an extra 
	drag due to induced viscous shear in the surrounding fluid~\cite{PGSMD75}.  
	In this paper, although the in-membrane flow is modified by membrane 
	deformation, any changes to surrounding flows can be neglected so long as 
	the characteristic length scale associated with the membrane deformation 
	$\sqrt{\kappa/\sigma}$, is much less than the Saffman-Delbr\"{u}ck length 
	$\eta/\mu$, where $\eta$ and $\mu$ are the two- and three- dimensional 
	viscosities of the membrane and surrounding fluid, respectively.  Using 
	values taken from  \cite{FQ+13} ($\eta = 6\times 
	10^{-10}\,\mathrm{kg}\,\mathrm{s}^{-1}$, 
	$\mu=10^{-3}\,\mathrm{kg}\,\mathrm{m}^{-1}\,\mathrm{s}^{-1}$, and $\kappa = 
	20\,\kT$ at room temperature) we find that the role of the embedding fluid 
	can be neglected for tensions $\sigma\gtrsim 2\times 
	10^{-7}\,\mathrm{N}\,\mathrm{m}^{-1}$, consistent with the full range of 
	investigated in \cite{FQ+13}.}.  The force balance condition for this 
	motion is then $F=-\lambda V$, where $F$ is the hydrodynamic stress 
	integrated over the walls of the protein \cite{Batchelor} and the sign 
	signifies that drag forces act opposite to the direction of motion.

This otherwise straightforward calculation is greatly complicated by the shape 
of the membrane, and requires the use of differential geometry.  For the 
uninitiated, a summary of both notation and relevant results is given in the 
SI.  In brief, at each point on the manifold, the components $\Pi^{ij}$ 
($i,j=1,2$) of the rank-(2,0) Cauchy stress tensor are defined with respect to 
a non-normalised basis $\boldsymbol{e}_i$, which spans the tangent plane to 
$\mathcal{S}$ at that point.  In order to calculate such stresses, both the 
hydrostatic pressure $p$ and components of the fluid velocity field $v^i$, are 
required, {\it i.e.},
\begin{equation}
	\Pi^{ij} = - p g^{ij} + \eta\left( v^{i;j} + v^{j;i}\right),
	\label{eq:Pi}
\end{equation}
where the constant $\eta$ is a two-dimensional viscosity and $g^{ij}$ are the 
components of the inverse metric.  Here, a comma ``,'' and semi-colon ``;'' 
placed before a lower index represent partial and covariant differentiation, 
respectively, whilst upper-indices may be lowered and lower-indices raised by 
contraction with the metric and its inverse, respectively ({\it i.e.}, 
$v^i=v_j\, g^{ij}$ and $v^{i;j} = {v^i}_{;k}\, g^{kj}$, {\it etc}.).  If the 
direction of motion of the inclusion is assumed (without loss of generality) to 
be in the $x$-direction, the net force $F$ becomes
\begin{equation}
	F = \int_{\partial\mathcal{S}} 
	\left(\hat{\boldsymbol{i}}\cdot\boldsymbol{e}_i\right)\Pi^{ij}\,dl_j
	=-\lambda V,
	\label{eq:drag_2}
\end{equation}
where $\partial\mathcal{S}$ is the boundary between the surface and the 
protein, and $dl_j$ is shorthand for the appropriate line 1-form(s).  Under 
steady state conditions, the hydrostatic pressure and fluid velocity fields 
satisfy the \textit{covariant} form of Stokes' 
equation~\cite{MLH+07,MLHAJL10,MAAD09}:
\begin{equation}
	\eta\left( {{v^{i}}_{;j}}^{;j} + K v^i\right) - p^{,i} = 0.
	\label{eq:Stokes_component}
\end{equation}
Here, the crucial difference with standard (Euclidean) hydrodynamics is that, 
if the membrane has a non-zero Gaussian curvature $K$, the shear stresses 
exerted by the fluid are modified.

In principle, the two equations (\ref{eq:Stokes_component}) can be solved, 
subject to boundary conditions, when combined with the constraint of 
incompressibility, ${v^i}_{;i} = 0$.  In practice, it is often easier to solve 
for a scalar stream function $\psi$ by writing
\begin{equation}
	v^i = \frac{1}{\sqrt{\left\vert g\right\vert}} \varepsilon^{ij} \psi_{,j},
	\label{eq:v_stream}
\end{equation}
where $\varepsilon^{ij}$ is a two-dimensional anti-symmetric Levi-Civita 
symbol, and $\vert g\vert$ is the determinant of the metric $g_{ij}$.  
Consigning the cumbersome derivation to the SI, we present the result in 
index-free notation using angle brackets $\langle\cdot,\cdot\rangle$ to 
indicate an inner product taken with respect to the metric
\begin{equation}
	\left(\frac{1}{2}\Delta + K\right)\Delta\psi + \langle\nabla 
	K,\nabla\psi\rangle = 0.
	\label{eq:psi2}
\end{equation}
Here, $\nabla$ is the gradient operator, extended to apply on a smooth 
manifold, and $\Delta$ is the Laplace-Beltrami operator.  
Equation~(\ref{eq:psi2}) is a fourth order partial differential equation which 
encapsulates incompressible Stokes flow on a two-dimensional smooth manifold 
(surface) in one single equation.  Notice that if the manifold is planar, {\it 
i.e.}, the Gaussian curvature is zero, then the usual biharmonic equation, 
$\Delta^2 \psi=0$, is recovered.

Unfortunately, for most non-trivial geometries, finding a closed-form solution 
to (\ref{eq:psi2}) is problematic.  However, approximate solutions may be 
constructed by considering the equation perturbatively.  In our case, both the 
Laplace-Beltrami operator and the Gaussian curvature may be expanded as power 
series in terms of the small angle $\alpha$.  We further postulate (and later 
verify) that $\psi$ can be expanded in the same way, {\it i.e.}, $\psi = 
\psi^{(0)} + \alpha\,\psi^{(1)}+\alpha^2\psi^{(2)} +O\left( \alpha^3 \right)$.  
Equation~(\ref{eq:psi2}) can now be solved order by order, subject to boundary 
conditions.  We impose a no-slip condition at the interface between the protein 
and the membrane, whilst as $r\to\infty$, we follow \cite{PGS75} and match with 
the leading term, in $r$, of a different velocity field, found by solving a 
Stokes equation that incorporates the extra drag from the embedding fluid.  At 
both boundaries, these conditions are satisfied at lowest order, leading to the 
following results.

\begin{figure}[t!]
\centering
\includegraphics[width=0.35\textwidth]{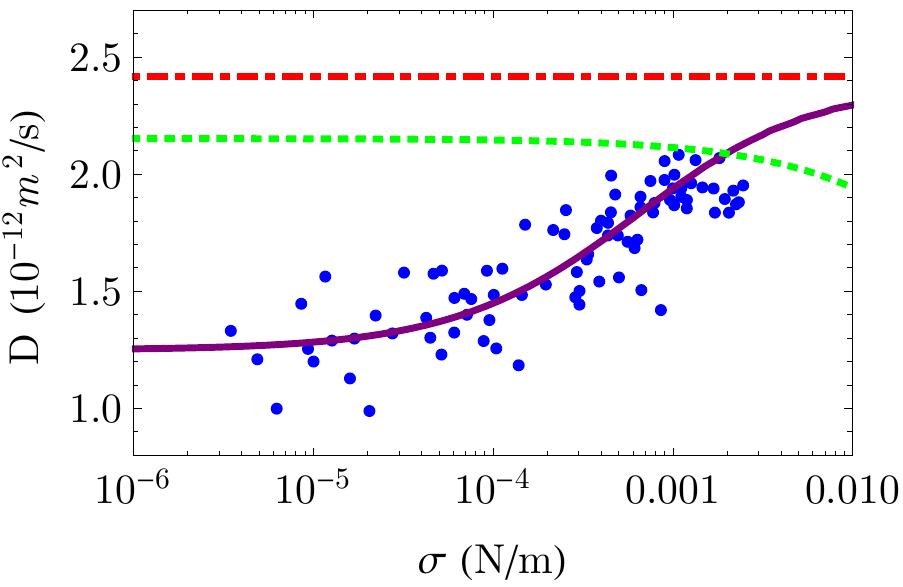}
\caption
{
	{\bf Tension-dependent diffusion.}  Log-linear plot of membrane tension 
	against diffusion constant for KvAP.  Blue points represent experimental 
	data from \cite{FQ+13}.  The red dot-dashed line is the tension independent 
	diffusion constant of a cylindrical inclusion of equivalent radius 
	\cite{PGS75}.  This emerges at zeroth order in our perturbative scheme for 
	KvAP.  The next lowest non-zero corrections must be calculated 
	numerically and are of order $\alpha^2$.  For proteins with a completely 
	rigid conformation (constant contact angle $\alpha=0.16$ Rad, irrespective 
	of tension) the hydrodynamic picture is not compatible with the data (green 
	dotted line).  However, if the 
	protein is permitted to deform elastically in response to the torque it 
	experiences on its walls we obtain an excellent single-parameter fit (solid purple line).
  	In all cases, the protein 
	radius $a=5$ nm, the membrane and solvent viscosities are $\eta = 6\times 
	10^{-10}$ kg s$^{-1}$ and $\mu=10^{-3}$ kg m$^{-1}$ s$^{-1}$ respectively 
	and the membrane rigidity is $\kappa = 20\,\kT$ at room temperature.
}
\label{fig:overlay}
\end{figure}

At lowest order, $\psi^{(0)}$ satisfies the biharmonic equation and the results 
of Saffman \cite{PGS75} are reproduced by design.  The resulting drag is 
$\lambda^{(0)} = 4\,\pi\,\eta/\mathcal{C}$, where $\mathcal{C}=\log (\eta/a\mu) 
- \gamma$, and $\gamma$ is Euler's constant.

At first order, $\psi^{(1)}$ also satisfies the biharmonic equation.  However,  
applying the boundary conditions gives $\psi^{(1)}=0$, implying that 
$\lambda^{(1)}=0$
\footnote
{
	There is a first order correction to the velocity 
	$\boldsymbol{v}=v^i\,\boldsymbol{e}_i$, because of the $\alpha$-dependence 
	of the basis vector $\boldsymbol{e}_1$.  However, this correction plays no 
	role in the lateral net force $F$, which acts in-plane by definition, and 
	therefore $\lambda^{(1)}=0$.
}.  This is a natural consequence of the up/down symmetry of the membrane: 
corrections to the drag coefficient $\lambda$ must be invariant under 
$\alpha\rightarrow-\alpha$.

At second order, $\psi^{(2)}$ satisfies an {\it inhomogeneous} biharmonic 
equation.  The general solution can be constructed by combining the solution to 
the homogeneous equation with a particular solution that can be calculated via 
an appropriate Green's function, see SI for details.  The resulting integrals 
must be calculated numerically \cite{Wolfram} and there is therefore no 
closed-form solution for $\lambda^{(2)}$.  Nevertheless, our result may still 
be compared with experiments~\cite{FQ+13} by invoking the Stokes-Einstein 
relation
\begin{equation}
	D= D^{(0)}\left[ 1 - \alpha^2\left(\lambda^{(2)}/\lambda^{(0)}\right) 
	\right] + O\left( \alpha^3 \right),
	\label{eq:D}
\end{equation}
where $D^{(0)} = \kT /\lambda^{(0)}=\kT\,\mathcal{C}/4\,\pi\,\eta$ is the 
diffusion coefficient of a cylindrical protein moving in a planar membrane 
\cite{PGS75}.  Here $\lambda^{(2)}$ depends implicitly on $\sigma$ through the 
shape of the membrane and hence the metric. Fig.~\ref{fig:overlay}
shows this result as a function of applied tension (green dotted curve).  By 
kind permission of the authors, our results are shown against the original data 
from \cite{FQ+13}.  We see that rigid proteins, assumed to have a constant 
contact angle $\alpha$, would experience a {\it reduction} in their diffusion 
constant at high tensions.  The reason is that the dimple induced in the 
membrane becomes an increasingly localised region of high Gaussian curvature, 
resulting in extra shear stresses in the fluid and hence extra drag on the 
protein. This indicates that, regardless of the tension, a completely rigid 
conical protein (otherwise resembling KvAP) will never diffuse like a 
cylindrical one, such as AQP0.

\begin{figure}[t!]
\centering
\includegraphics[width=0.35\textwidth]{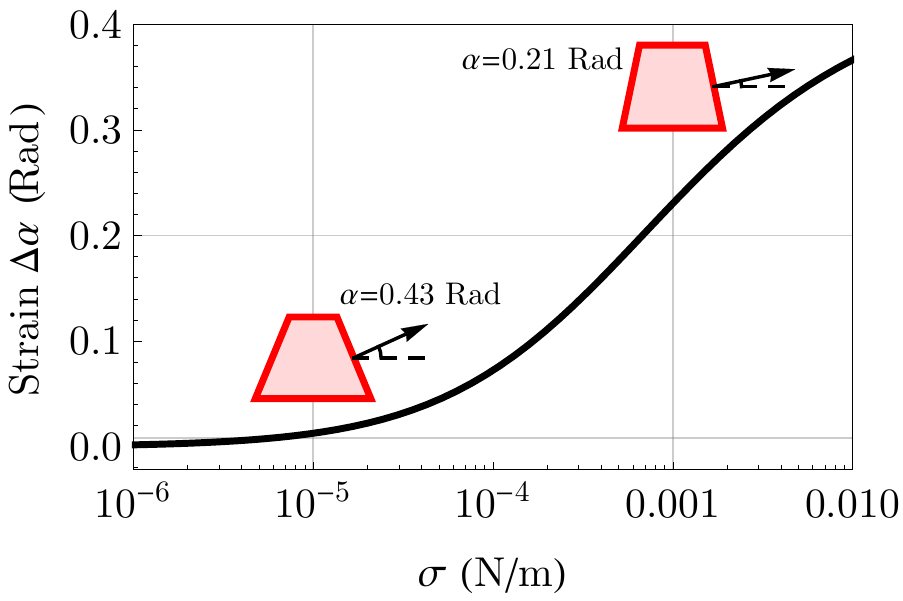}
\caption
{
	{\bf Protein shape changes.} Log-linear plot of angular strain against 
	membrane tension.  In the physiological range investigated by \cite{FQ+13}, 
	{\it i.e.,} $10^{-5}$---$10^{-3}$ N/m, we predict an angular variation of 
	around 0.22 Rad, roughly equivalent to a material strain of about 20\%.
}
\label{fig:strain}
\end{figure}

We therefore propose that the shape of the protein changes with tension, and 
invoke linear torsional response $\tau=\tau_r + k\left( \alpha-\alpha_r 
\right)$.  The torque $\tau$ exerted on the ``walls'' of the protein can be 
found from the boundary terms in the earlier variational analysis
\begin{equation}
	\tau = 2\pi\,a\,\sigma\,h(a)\,\alpha.
\label{torque}
\end{equation}
The subscript $r$ denotes ``reference'', where $\tau_r$ is calculated by 
identifying the tension $\sigma_r$ at which the green dotted line of 
Fig.~\ref{fig:overlay} intersects the data, and then substituting both $\sigma 
= \sigma_r$ and $\alpha=\alpha_r=0.16$ Rad ({\it i.e.}, the angle used in 
\cite{FQ+13}) into Eq.~(\ref{torque}).  The result is a tension-dependent 
expression for the angle $\alpha(\sigma)$, which depends on the torsional 
stiffness $k$.  Using a least-squares procedure, a single parameter fit for $k$ 
gives excellent agreement with the data (solid purple line in 
Fig.~\ref{fig:schematic}) yielding a value of $k=26.8\,\kT$ at room 
temperature.  Reassuringly, this is entirely consistent with the energies 
required for voltage activation~\cite{KJS08}.  Moreover, we predict 
non-negligible angular strains $\Delta \alpha:= \alpha_0 - \alpha(\sigma)$, 
where $\alpha_0:=\lim_{\sigma\to 0} \alpha(\sigma)=0.44$ Rad, for the range of 
tensions investigated in \cite{FQ+13}, see Fig.~\ref{fig:strain}.

In the context of our evidence for significant structural strains at 
physiological tensions, a reassessment of the function and structure of 
membrane proteins under tension may be required.  Our results are especially 
pertinent since the highly specialised functions of membrane-embedded proteins 
are currently thought to require precise spatial positioning of at least the 
key functional residues \cite{MacKinnon,McCoy}.  We therefore welcome further 
work in the area.


\section{Acknowledgments}
\begin{acknowledgments}
We thank the authors of \cite{FQ+13} for discussions prior to publication and 
for permission to use their data. We acknowledge discussions with G.~Rowlands 
(Warwick), A.~Rautu (Warwick) and P.~Sens (Paris) as well as discussion and 
critical reading of the manuscript by D.~R.~Daniels (Swansea).  We also 
acknowledge EPSRC grant EP/I005439/1 (a Leadership Fellowship to MST).
\end{acknowledgments}

\section*{SUPPLEMENTARY MATERIAL}

This section is intended to provide both theoretical background (for example, 
in differential geometry) and also certain calculation details (such as the 
second order coefficient $\psi^{(2)}$ of the perturbative expansion of the 
stream function) which, although interesting, are secondary to the message of 
the main manuscript.

\subsection*{Membrane shape}
\begin{figure}
\centering
\includegraphics[width=0.35\textwidth]{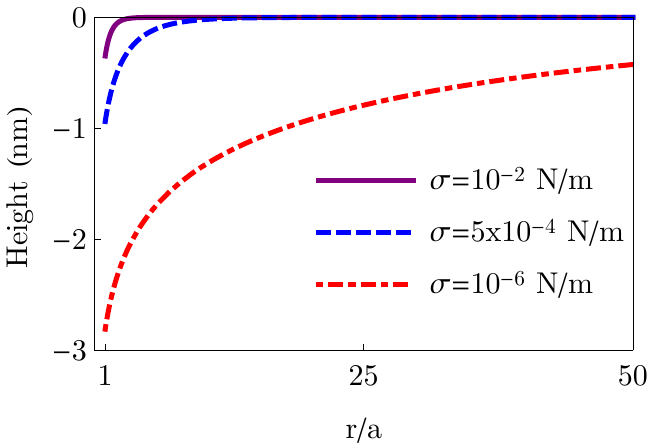}
\caption
{
	\textbf{The effect of changing tension.}  Plot of Eq.~(\ref{sm:h}), the 
	height profile of the membrane, as a function of the (non-dimensional) 
	radial coordinate $r/a$ ($a=5$ nm, $\alpha=0.16$ Rad, 
	$\kappa=8.2\times10^{-20}$ J).  For the range of tensions investigated in 
	\cite{FQ+13}, the localisation of the ``dimple'' caused by KvAP varies 
	significantly.
}
\label{fig:tension}
\end{figure}
Associating a smooth Reimannian manifold, or regular surface, $\mathcal{S}$, 
with the mid-plane of the bi-layer, the shape induced by KvAP can be calculated 
by using a variational approach.  Following the main text, the total 
(Helmholtz) free energy of the membrane is
\begin{equation}
	E := \int_\mathcal{S} \left[ \frac{\kappa}{2} (2H)^2 + \sigma\right]dA,
	\label{sm:helmholtz}
\end{equation}
where $dA$ is used as a shorthand for the volume 2-form, $\mathrm{vol}^2$, 
associated with the manifold $\mathcal{S}$ (see next Section).
The shape of the membrane at equilibrium is found by minimising 
(\ref{sm:helmholtz}) over a family of surfaces.  Here, since the 
parameterisation must be radially symmetric, a polar Monge approach will 
suffice.  That is, each surface is characterised by a height field 
$\epsilon\,h(r)$, where $r\in [a,\infty)$.  The distance $h$ is measured in the 
	direction normal to the plane and $\epsilon$ is a small number to help 
	ensure that, formally, we are restricted to single-valued surfaces.  
	Expanding in powers of $\epsilon$ and setting the variation equal to zero 
	leads to a fourth order Euler-Lagrange equation in one-dimension:
\begin{equation}
	\left[ \frac{1}{r}\frac{d}{dr}
	\left(r\frac{d}{dr}\right) - \frac{\sigma}{\kappa}
	\right]
	\left[\frac{1}{r}\frac{d}{dr}
		\left(r\frac{d}{dr}\right)\right]h(r) = 0.
	\label{sm:E-L}
\end{equation}
At large $r$, the boundary conditions are that the height function and its 
first derivative must vanish, {\it i.e.} $\lim_{r\to\infty} h(r) = 0$, and 
$\lim_{r\to\infty} d\,h / d r =0$.  At the inclusion, we have $\left.\epsilon\, 
(d\,h/dr)\right\vert_{r=a} = -\tan{\alpha}\simeq -\alpha$, where $\alpha$ is 
the small angle subtended at the inclusion.  The solution to (\ref{sm:E-L}) can 
be shown to be \cite{TRW+98}
\begin{equation}
	\epsilon\, h(r) = \alpha\,l\, K_0 \left( r/l \right)\,/\,K_1 \left( 
	a/l\right),
	\label{sm:h}
\end{equation}
where $a$ is the radius of the protein, $l=\sqrt{\kappa/\sigma}$ is the 
membrane correlation length and $K_n$ is an order-$n$ modified Bessel function 
of the second kind.  Equation (\ref{sm:h}) implies that $\epsilon=\alpha$, 
giving a convenient interpretation for our small parameter as the angle, at the 
interface of the membrane and the inclusion, between the outward normal of the 
inclusion and the plane defined by the $z$-axis (see Fig.~2 of the main text).  
Moreover, plotting (\ref{sm:h}) for increasing values of surface tension 
indicates that the disturbance caused by the protein becomes increasingly 
localised (see Fig.~1).

\subsection*{Differential geometry}
Before considering Stokes' flow on a manifold, it is first necessary to 
understand the basic aspects and notation of differential geometry.  Here, we 
aim to provide only what is necessary or helpful to understand the main article 
and the focus is therefore two-dimensional Riemannian manifolds.

The components of a vector $\boldsymbol{v}$ and its corresponding 1-form $v$ 
are distinguished by superscript and subscript, respectively.  That is, if a 
two-dimensional smooth Riemannian manifold--- the surface--- is parameterised 
by coordinates $\{u^i:i=1,2\}$ then $\boldsymbol{v} = v^i \boldsymbol{e}_i$ and 
$v = v_i \mathrm{d}u^i$, where an implicit sum is understood by repeated 
indices of different type ({\it i.e.}, upper and lower).  Under a change of 
coordinates, the transformation properties of the functions $v^i$ and $v_i$ can 
be readily calculated.  The former is said to transform in a contravariant way, 
and the latter in a covariant way.  Here, if $\boldsymbol{R}(u^1,u^2)$ gives 
the position of points on the surface relative to some origin, then
\begin{equation}
	\boldsymbol{e}_i:=\frac{\partial \boldsymbol{R}(u^1,u^2)}{\partial u^i},
	\label{sm:e_alpha2}
\end{equation}
are basis vectors (not normalised) for the tangent space at each point on the 
surface.  Furthermore, $\{\mathrm{d}u^i: i=1,2\}$ form the corresponding basis 
of 1-forms, such that $\mathrm{d}u^i (\boldsymbol{e}_j) = \delta^i_j$, where 
$\delta^i_j$ is the Kronecker delta symbol.

Since the manifold is Riemannian, it is equipped with a positive-definite 
metric, whose components $g_{ij} := \boldsymbol{e}_i\cdot\boldsymbol{e}_j$, 
induce an inner product, which we denote by angle brackets 
$\langle\cdot,\cdot\rangle$.  Specifically, for arbitrary vectors (of the same 
dimension) $\boldsymbol{v}$ and $\boldsymbol{w}$, we may define 
$\langle\boldsymbol{v},\boldsymbol{w}\rangle := v^i g_{ij} w^j$.  The inner 
product then permits the explicit identification of vectors, {\it e.g.}, 
$\boldsymbol{v}$, with its dual 1-form, $v$, by the condition 
$v(\boldsymbol{w})=\langle \boldsymbol{v},\boldsymbol{w}\rangle$, which holds 
for all $\boldsymbol{w}$.  Noticing that $v(\boldsymbol{w}) = v_i \mathrm{d}u^i 
(\boldsymbol{w}) = v_i w^i$ and using the above definition of the inner product 
of two vectors implies the raising and lowering properties of the metric and 
its inverse ($g^{ij}$), respectively.  That is, $v_i = g_{ij} v^j$ and $v^i = 
g^{ij} v_j$.  Using this property, the inner product acting on two 1-forms can 
be defined in a complementary way to that of the inner product on vectors:
\begin{equation}
	\langle v,w\rangle := v_i g^{ij} w_j = v_i w^i= 
	\langle\boldsymbol{v},\boldsymbol{w}\rangle.
	\label{sm:inner_1_form}
\end{equation}
The identification of vectors with 1-forms may also be used to describe the 
action of the exterior derivative operator $\mathrm{d}$ on a scalar field 
(0-form) $\phi$.  That is, $\mathrm{d}\phi(\boldsymbol{v}) = \langle \nabla 
\phi,\boldsymbol{v}\rangle$ for all vectors $\boldsymbol{v}$, where
\begin{equation}
	\nabla \phi := g^{ij}\frac{\partial \phi}{\partial u^i} \boldsymbol{e}_j,
	\label{sm:grad}
\end{equation}
is just the usual gradient operator, extended to apply on a smooth manifold.  
In general, the exterior derivative takes a $k$-form to $(k+1)$-form, though 
its full definition is not required here.  Nevertheless, it is helpful to list 
some properties of $\mathrm{d}$.  First, repeated application always yields 
zero.  That is, $\mathrm{d}\mathrm{d}v=\mathrm{d}^2 v=0$ for an arbitrary 
differential form $w$.  Second, the action of $\mathrm{d}$ is distributive over 
the wedge product.  For example, $\mathrm{d}\left(v\wedge w\right) = 
\mathrm{d}v\wedge w - v\wedge \mathrm{d}w$, where $v$ and $w$ are 1-forms.  
Here, the wedge product is just an anti-symmetrised tensor product $v\wedge w = 
v\otimes w - w\otimes v = -w\wedge v$, which is very natural in geometrical 
systems.  In particular, the volume 2-form for a two-dimensional surface is 
written as the following wedge product
\begin{equation}
	\mathrm{vol}^2 = \sqrt{\vert g\vert}\,\mathrm{d}u^1 \wedge\mathrm{d}u^2,
	\label{sm:vol^2}
\end{equation}
where $\vert g\vert:=\det g_{ij}$.  The volume form in two-dimensions is an 
area, and often written using the shorthand ``$dA$''.  It can be used to define 
a pair of line 1-forms ``$dl_i$'', which are natural in the given coordinate 
system:
\begin{equation}
	dl_i:=\mathrm{vol}^2\left(\boldsymbol{e}_i\right) = \sqrt{\left\vert 
	g\right\vert}\, \sum_{j>i} \varepsilon_{ij}\, \mathrm{d}u^j,
	\label{sm:line}
\end{equation}
where $\varepsilon_{ij}$ is a two-dimensional Levi-Civita tensor density of 
weight $-1$.  Notice that the sum only permits values of $i>j$, therefore 
$dl_1=\sqrt{\left\vert g\right\vert} \mathrm{d}u^2$ and $dl_2=0$.

The covariant derivative (or Levi-Civita connection) at a point $x$ is an 
extension of the directional derivative.  It takes two arguments: a direction 
vector $\boldsymbol{T}$ defined in the tangent plane at $x$ and a tensor field 
over the tangent bundle to the manifold, that must be smoothly varying in the 
neighbourhood of $x$.  Assuming a two-dimensional smooth (Riemannian) manifold, 
with coordinates $u^i$, the action on a scalar field $\phi$ is then 
$\nabla_{\boldsymbol{T}} \phi := \phi_{,j}\,\mathrm{d}u^j (\boldsymbol{T})$, 
where a subscript comma ``,'' is shorthand for a partial derivative, {\it 
i.e.},
\begin{equation}
	{\phi}_{,j} := \frac{\partial \phi}{\partial u^j}.
	\label{sm:comma_scalar}
\end{equation}
When acting on a vector $\boldsymbol{v} = v^i \boldsymbol{e}_i$, we write 
$\nabla_{\boldsymbol{T}} \boldsymbol{v} := \boldsymbol{e}_i \left( 
{v^i}_{;j}\right)\mathrm{d}u^j (\boldsymbol{T})$, where the components 
${v^i}_{;j}$ are given by
\begin{equation}
	{v^i}_{;j} := {v^i}_{,j} + v^k \Gamma^i_{jk}.
	\label{sm:cov_components}
\end{equation}
Once again, a subscript comma ``,'' is shorthand for a partial derivative,
\begin{equation}
	{v^i}_{,j} := \frac{\partial v^i}{\partial u^j},
	\label{sm:comma}
\end{equation}
whilst the $\Gamma^i_{jk} = g^{ip}\left(g_{pj,k} + g_{pk,j} - 
g_{jk,p}\right)/2$ are Christoffel symbols, which define the action of the 
covariant derivative, via $\nabla_{\boldsymbol{e}_i} \boldsymbol{e}_j = 
\boldsymbol{e}_k \Gamma^k_{ij}$.  Note that the shorthand $\nabla_i := 
\nabla_{\boldsymbol{e}_i}$ is frequently used in physics.  For a $1$-from, the 
action of the covariant derivative can be defined by demanding that the 
``Leibniz rule'' holds.  That is, if a scalar field is defined by the action of 
a $1$-form on a vector, {\it i.e.}, $\phi:=v(\boldsymbol{w})=v^i\,w_i$, then
\begin{equation}
	\nabla_i \left(v^j\,w_j\right) = \left(v^j\,w_j\right)_i := {v^j}_{;i}\, 
	w_j + v^j \,w_{j;i}.
	\label{sm:Leibnitz}
\end{equation}
The result is that
\begin{equation}
	v_{i;j}:= v_{i,j} - v_k\,\Gamma^k_{ij},
	\label{sm:cov_1form}
\end{equation}
which is consistent with the notion of using the metric as a raising / lowering 
operator ({\it i.e.}, $v_{i;j}=g_{ij}{v^j}_{;k}$).

\subsection*{Stokes' flow on a manifold}
As described in the main text, the aim is to calculate the drag coefficient 
associated with an inclusion moving though a two-dimensional fluid.  If the 
protein is moving with a steady velocity $\boldsymbol{V} = V 
\hat{\boldsymbol{i}}$ then the force balance equation for hydrodynamic drag is 
given by
\begin{equation}
	F = \int_{\partial\mathcal{S}} 
	\left(\hat{\boldsymbol{i}}\cdot\boldsymbol{e}_i\right)\Pi^{ij}\,dl_j
	=-\lambda V,
	\label{sm:drag_2}
\end{equation}
where $\partial\mathcal{S}$ is the boundary to the surface at the inclusion and 
$dl_j$ is shorthand for the line 1-form.  The Cauchy stress tensor is a 
rank-$(2,0)$ tensor, with components given by
\begin{equation}
	\Pi^{ij} := - p g^{ij} + \eta\left( v^{i;j} + v^{j;i}\right),
	\label{sm:Pi}
\end{equation}
where $p$ is the hydrostatic pressure field, the constant $\eta$ is a 
two-dimensional viscosity.  So, in order to calculate the drag, it is first 
necessary to know both the pressure and velocity fields, {\it i.e.}, to solve 
Stokes' equation.  As mentioned in the main text, the derivation of Stokes' 
equation on a manifold already exists in the literature.  The result, in 
component form, is that
\begin{equation}
	\eta\left( {{v^{i}}_{;j}}^{;j} + K v^i\right) - p^{,i} = 0.
	\label{sm:Stokes_component}
\end{equation}
In principle, (\ref{sm:Stokes_component}) can be solved, for given boundary 
conditions, when taken together with the condition of incompressibility,
\begin{equation}
	{v^i}_{;i} = 0.
	\label{sm:incompress_component}
\end{equation}
In practice, it is often easier to use a stream function $\psi$, defined such 
that
\begin{equation}
	v^i = \frac{1}{\sqrt{\left\vert g\right\vert}} \varepsilon^{ij} \psi_{,j},
	\label{sm:v_stream}
\end{equation}
where $\varepsilon^{ij}$ is a two-dimensional anti-symmetric Levi-Civita 
symbol, and $\vert g\vert$ is the determinant of the metric $g_{ij}$.  We 
remark that, by taking the embedding space into account, 
Eq.~(\ref{sm:v_stream}) is equivalent to writing $\boldsymbol{v} = 
\hat{\boldsymbol{n}}\times\nabla\psi$, where $\hat{\boldsymbol{n}}$ is the unit 
normal to the surface and $\nabla$ is the gradient operator of the manifold 
[{\it cf}. Eq.~(\ref{sm:grad})].  Furthermore, by definition, taking the 
divergence of (\ref{sm:v_stream}) by applying the covariant derivative and 
contracting over the two free indices, yields zero.

At this point, we wish to derive an equation that, given boundary conditions, 
can be used to calculate $\psi$.  We start by noticing that, in fact, under 
coordinate transformation, $\varepsilon^{ij}$ transforms as a tensor density of 
weight $+1$ (and $\varepsilon_{ij}$ with a weight $-1$).  This means that 
$\varepsilon^{ij}/\sqrt{\left\vert g\right\vert}$ behaves like a (pure) tensor, 
and hence gives Eq.~(\ref{sm:v_stream}) the correct transformation properties.  
It is clear that
\begin{equation}
	\frac{1}{\sqrt{\left\vert g\right\vert}} \varepsilon^{ij} = \left\{
		\begin{array}{ll}
			1/\sqrt{\left\vert g\right\vert} & \mathrm{for}\ (i,j)=(1,2)\\
			-1/\sqrt{\left\vert g\right\vert} & \mathrm{for}\ (i,j)=(2,1)\\
			0 & \mathrm{for}\ i=j
		\end{array}
\right. ,
	\label{sm:epsilon}
\end{equation}
which is helpful when substituting (\ref{sm:v_stream}) into 
(\ref{sm:Stokes_component}).  By using the distributive property of the 
covariant derivative--- {\it i.e.}, $(\phi\, a^i)_{;j} = \phi_{,j}\, a^i + 
\phi\, {a^i}_{;j}$ for scalar field $\phi$ and vector components $a^i$---  
together with the fact that the covariant derivative of the determinant of the 
metric vanishes, we see that (\ref{sm:Stokes_component}) becomes
\begin{equation}
	\frac{\eta\,\varepsilon^{ij}}{\sqrt{\left\vert g\right\vert}}\left[ 
	{\psi_{,j;k} }^{;k} + K\,\psi_{,j} \right] - p^{,i}=0.
	\label{sm:pre_permute}
\end{equation}
The first term can be further simplified by the invoking the rules for 
commuting covariant derivatives.  In order to see this, it is first helpful to 
make the contraction over the index ``$k$'' explicit, by writing it as the 
action of a Kronecker delta, {\it i.e.}, ${\psi_{,j;k} 
}^{;k}=\delta^k_l\,{\psi_{,j;k} }^{;l}$.  From here, we note that 
$\psi_{,j;k}=\psi_{,k;j}$, which can be seen from the definition 
(\ref{sm:cov_1form}).  Next, we may apply the commutation relation
\begin{equation}
	{\psi_{,k;j}}^{;l} = {{\psi_{,k}}^{;l}}_{;j} - 
	g^{lp}\,\mathcal{R}^q_{kpj}\,\psi_{,q},
	\label{sm:commute}
\end{equation}
where $\mathcal{R}^q_{kpj}$ are the components of the Riemann tensor, which is 
given in two-dimensions by the Bianchi identity $\mathcal{R}^q_{kpj} = K\, 
g^{ql}\left(g_{lp}\,g_{kj}-g_{lj}\,g_{kp}\right)$.  Re-applying the contraction 
implied by $\delta^k_l$, the result is that
\begin{equation}
	\frac{\eta\,\varepsilon^{ij}}{\sqrt{\left\vert g\right\vert}}\left[ 
	\left(\Delta \psi\right)_{,j} + 2K\,\psi_{,j} \right] - p^{,i}=0.
	\label{sm:post_permute}
\end{equation}
Here, $\Delta$ is the Laplace-Beltrami operator
\begin{equation}
	\Delta := \frac{1}{\sqrt{\vert g\vert}}\frac{\partial}{\partial 
	u^i}\left(\sqrt{\vert g\vert} g^{ij} \frac{\partial}{\partial 
	u^j}\right),
	\label{sm:Lap-Bel}
\end{equation}
defined such that $\Delta\psi = {\psi_{,i}}^{;i}$.  Finally, the pressure term 
in this equation may be eliminated by taking the two-dimensional analog of the 
curl ($\mathrm{curl}\,\boldsymbol{a} = \varepsilon_{ij} a^{j;i}$).  Using angle 
brackets $\langle\cdot,\cdot\rangle$ according to (\ref{sm:inner_1_form}) the 
resulting equation can be written in index-free notation as
\begin{equation}
	\left(\frac{1}{2}\Delta + K\right)\Delta\psi + \langle\nabla 
	K,\nabla\psi\rangle = 0.
	\label{sm:psi2}
\end{equation}
As remarked in the main text, for most non-trivial geometries, finding a 
closed-form solution to (\ref{sm:psi2}) is problematic.  Therefore, in this 
paper, we find approximate solutions by considering the equation 
perturbatively.

\subsection*{Perturbation theory}
In order to treat (\ref{sm:psi2}) perturbatively, both the Laplace-Beltrami 
operator and the Gaussian curvature must be expanded as power series in terms 
of $\alpha$, then, postulating that $\psi$ can be expanded in the same way,
\begin{equation}
	\psi = \psi^{(0)} + \alpha\psi^{(1)}+\alpha^2\psi^{(2)} +O\left( \alpha^3 
	\right),
	\label{sm:psi_pert}
\end{equation}
Eq.~(\ref{sm:psi2}) may be solved order by order.  As in the main text, 
bracketed superscripts indicate the order, in $\alpha$, of each term.  That is, 
$\psi^{(0)}$ is of order constant ({\it i.e.}, $\alpha^0=1$) and $\psi^{(1)}$ 
is order $\alpha$, and so on and so forth.  In order to understand these 
corrections, recall that the control parameter $\alpha$ has a clear 
interpretation as an angle, and therefore the perturbation theory is best 
though of geometrically.

As indicated in Fig.~2 of the main text, polar coordinates $r$ and $\theta$ are 
used, with the origin on the axis of symmetry of the inclusion.  Each point on 
the surface then has position relative to the origin of 
$\boldsymbol{R}(r,\theta)=r\hat{\boldsymbol{r}} - \alpha 
h(r)\hat{\boldsymbol{z}}$, where $\hat{\boldsymbol{r}}$ and 
$\hat{\boldsymbol{z}}$ are just the usual vectors in cylindrical polars, and 
$h(r)$ is given by (\ref{sm:h}).  The definition (\ref{sm:e_alpha2}) implies 
that the tangent vectors at each point on the surface are then just
\begin{equation}
	\boldsymbol{e}_1:=\frac{\partial \boldsymbol{R}}{\partial r} = 
	\hat{\boldsymbol{r}}- \alpha h'
	\hat{\boldsymbol{z}},\ \mathrm{and}\ \boldsymbol{e}_2:=\frac{\partial 
		\boldsymbol{R}}{\partial \theta} =
	r\hat{\boldsymbol{\theta}},
	\label{sm:tangent_vecs}
\end{equation}
where a dash is used as shorthand for derivative, {\it i.e.}, $h':=dh/dr$.  
(Recall that the vectors $\{\boldsymbol{e}_i : i=1,2\}$ are not normalised).
Using these definitions, the components of the metric
\begin{equation}
	g_{ij} = \left(
		\begin{array}{cc}
			1 + \alpha^2\left( h' \right)^2 & 0 \\
			0 & r^2
		\end{array}
		\right),
\end{equation}
and its inverse
\begin{equation}
		g^{ij} = \left(
		\begin{array}{cc}
			\left[1 + \alpha^2\left( h' \right)^2\right]^{-1} & 0 \\
			0 & 1/r^2
		\end{array}
		\right),
\end{equation}
can be calculated, from which it is immediately clear that $\vert 
g\vert=r^2\left[1+\alpha^2\left( h'\right)^2\right]$. Furthermore, the 
Christoffel symbols may be computed as
\begin{equation}
	\Gamma^1_{ij} = \left(
		\begin{array}{cc}
			\alpha^2\,h'\,h''\left[1 +\alpha^2\left( h' \right)^2\right]^{-1}  
			& 0 \\
		0 & r\left[1 + \alpha^2\left( h' \right)^2\right]^{-1}\end{array}
	\right),
\label{sm:Gamma_pert_1}
\end{equation}
and
\begin{equation}
	\Gamma^2_{ij} = \left(
		\begin{array}{cc}
			0 & 1/r \\
			1/r & 0
		\end{array}
		\right).
		\label{sm:Gamma_pert_2}
\end{equation}
Note that the $\Gamma_{ij}^k$ are not required for calculating either $\psi$ or 
the component functions $v^i$ [via Eq.~(\ref{sm:v_stream})].  However, they are 
necessary when computing covariant derivatives of $v^i$, which appear in the 
definition of the stress tensor $\Pi^{ij}$ [{\it cf.} Eq.~(\ref{sm:Pi})].

Using the above results and the definition (\ref{sm:Lap-Bel}), we see that the 
Laplace-Beltrami operator may be expanded in the following way
\begin{equation}
	\begin{split}
		\Delta &= \frac{1}{r}\frac{\partial}{\partial 
		r}\left(r\frac{\partial}{\partial r}\right) + 
		\frac{1}{r^2}\frac{\partial^2}{\partial \theta^2}
		-\alpha^2 \frac{h'}{r}\frac{\partial}{\partial 
		r}\left(rh'\frac{\partial}{\partial r}\right) + 
		O\left(\alpha^3\right)\\
		&=\Delta^{(0)} + \alpha^2\Delta^{(2)} + O\left(\alpha^3\right).
	\end{split}
	\label{sm:Lap-Bel_pert}
\end{equation}
Here, notice that the second order correction is an operator that acts on the 
radial variable only, as expected on symmetry grounds.  In order to calculate 
the Gaussian curvature, we invoke the Brioschi formula which, since the metric 
is diagonal, simplifies greatly, leading to
\begin{equation}
	\begin{split}
	K &= -\frac{1}{2\sqrt{\vert g\vert}}\left[\frac{\partial}{\partial 
	r}\left(\frac{g_{11}}{\sqrt{\vert g\vert}}\right) + 
\frac{\partial}{\partial \theta}\left(\frac{g_{22}}{\sqrt{\vert 
g\vert}}\right)\right]\\
&= \alpha^2\frac{h'h''}{r} + O\left(\alpha^3\right)\\
&=\alpha^2K^{(2)}+ O\left(\alpha^3\right).
\end{split}
	\label{sm:K_pert}
\end{equation}
In summary, the lowest order corrections to both the Laplace-Beltrami operator 
and the Gaussian curvature occur at second order [and the $O(1)$ contribution 
to the Gaussian curvature is zero].  Along with a set of given boundary 
conditions, the above formulas are all that is necessary to solve for the 
stream function $\psi$ order by order.

\subsection*{Boundary conditions}
The inclusion is moving with a constant velocity $V$, therefore imposing a 
no-slip condition at the interface between the inclusion and the membrane, 
gives
\begin{equation}
	v^1 (a,\theta)= V\cos\theta,\ \mathrm{and}\ v^2 (a,\theta)= 
	-\frac{V}{a}\sin\theta,
	\label{sm:bound_v}
\end{equation}
where the in-plane angle $\theta$ is measured from the positive $x$-direction.  
The conditions (\ref{sm:bound_v}) are true to all orders of $\alpha$, however, 
in our case, they are fully satisfied at lowest order, {\it i.e.},
\begin{equation}
	\left[ v^1 \right]^{(0)} (a,\theta)= V\cos\theta,\ \mathrm{and}\ \left[v^2 
	\right]^{(0)} (a,\theta)= -\frac{V}{a}\sin\theta,
	\label{sm:bound_v^0}
\end{equation}
where bracketed superscripts are once again used to indicate coefficients in a 
series expansion in terms of $\alpha$:
\begin{equation}
	v^i = \left[ v^i \right]^{(0)} + \alpha\left[ v^i 
	\right]^{(1)}+\alpha^2\left[ v^i \right]^{(2)} +O\left( \alpha^3 \right).
	\label{sm:v^i_pert}
\end{equation}
The result of the above is that all but the lowest order coefficients of this 
expansion must therefore vanish at the boundary.  That is
\begin{equation}
	\left[ v^i \right]^{(n)}(a,\theta)=0\ \forall\ n\in\mathbb{Z}_+.
	\label{sm:bound_v^n}
\end{equation}
As mentioned earlier, rather than use these conditions to calculate the 
functions $v^i$ directly, it is easier to first solve (\ref{sm:psi2}) for the 
stream function and then use Eq.~(\ref{sm:v_stream}).  To this end, we must 
translate the boundary conditions on the component functions of the velocity 
field into those for the function $\psi$.  In general [see 
Eq.~(\ref{sm:v_stream})] we have the following order-by-order relationships
\begin{equation}
	\left[ v^i \right]^{(0)} = \frac{\varepsilon^{ij}}{r} \psi^{(0)}_{,j},\ 
	\left[ v^i \right]^{(1)} = \frac{\varepsilon^{ij}}{r} \psi^{(1)}_{,j},\ 
	\label{sm:v_stream_0_1}
\end{equation}
and
\begin{equation}
	\left[ v^i \right]^{(2)} = \frac{\varepsilon^{ij}}{r} \left[ 
	\psi^{(2)}_{,j} - \frac{\left( h' \right)^2}{2}\psi^{(0)}_{,j}\right].  
	\label{sm:v_stream_2}
\end{equation}
At the boundary, using the definition of $h(r)$ and applying both 
(\ref{sm:bound_v^0}) and (\ref{sm:bound_v^n}) gives
\begin{equation}
	\psi^{(0)}(a,\theta)=V a \sin\theta,\ \psi^{(1)}(a,\theta)=0,\ 
	\label{sm:bound_psi^0^1}
\end{equation}
and
\begin{equation}
	\psi^{(2)}(a,\theta) = (Va / 2)\sin\theta.
	\label{sm:bound_psi^2}
\end{equation}
By contrast, as $r$ approaches infinity, we follow Saffman \cite{PGS75} and 
impose that
\begin{equation}
	\lim_{r\to\infty}\psi^{(0)}(r,\theta)=\frac{U r 
	\sin\theta}{\mathcal{C}}\left(\mathcal{C} + \frac{1}{2} - 
	\log\frac{r}{a}\right),
	\label{sm:lim_psi_infty}
\end{equation}
%
%
%
where $\mathcal{C} :=\log (\eta / a\mu ) - \gamma$ is a constant.  Here, 
$\gamma$ is Euler's constant, and $\mu$ is the (three-dimensional) viscosity of 
the embedding fluid.  The condition (\ref{sm:lim_psi_infty}) comes from a 
matching criterion that arises when solving a Stokes'-like equation that has 
been modified due to the drag that results from a no-slip condition with the 
embedding fluid \cite{PGS75}.  Once again, at all higher orders, we simply have
\begin{equation}
	\lim_{r\to\infty}\psi^{(n)}(r,\theta)=0,\ \forall\ n\in\mathbb{Z}_+.
	\label{sm:psi_infty}
\end{equation}

\subsection*{Separation of angular dependence}
An immediate consequence of the above boundary conditions is that the 
$\theta$-dependence of $\psi$ is trivial.  That is, by writing
\begin{equation}
	\begin{split}
	\psi(r,\theta) &= \phi(r)\sin\theta \\
	&= \left[\phi^{(0)}(r) + \alpha\phi^{(1)}(r) + \alpha^2\phi^{(2)}(r) + 
	O(\alpha^3)\right]\sin\theta,
\end{split}
	\label{sm:psi_seperable}
\end{equation}
Eq.~(\ref{sm:psi2}) can be reduced to a fourth order {\it ordinary} 
differential equation (as opposed to a partial differential equation). Up to 
$O(\alpha^2)$, this can be seen by noting that only the $O(1)$ part of the 
Laplace-Beltrami operator, $\Delta^{(0)}$, acts on the variable $\theta$.  
Higher order corrections, such as $\Delta^{(2)}$ and $K^{(2)}$, are concerned 
with the radial variable only.  Moreover, given the $\theta$-dependence of 
$\psi$, the second partial derivative
$\partial^2 / \partial \theta^2$ contained in the definition of $\Delta^{(0)}$ 
leads to the following relation: $\Delta^{(0)} \psi (r,\theta) =   
\hat{B}\left[ \phi(r)\right]\sin\theta$.  Here, $\hat{B}$ is then a 
second-order ordinary differential operator
\begin{equation}
	\hat{B}:=\frac{1}{r}\frac{d}{dr}\left(r\frac{d}{dr}\right) - \frac{1}{r},
	\label{sm:B_k}
\end{equation}
whose eigenfunctions are modified Bessel functions of order one.

\subsection*{Calculating the stream function}
With the variable dependencies separated according to (\ref{sm:psi_seperable}) 
we may proceed to solve (\ref{sm:psi2}) for $\phi$ (and hence $\psi$) order by 
order.

\subsubsection*{Zeroth order}
At lowest order, $\psi^{(0)}$ satisfies the biharmonic equation, which, in 
terms of the radial function $\phi^{(0)}$ translates to
\begin{equation}
	\begin{split}
		&\hspace{20mm}\hat{B}^2 \phi^{(0)}(r) = 0,\\
	&\Longrightarrow\ \phi^{(0)}(r) = \frac{\mathcal{C}_1}{r} + \mathcal{C}_2 r 
	+ \mathcal{C}_3 r^3 + \mathcal{C}_4 r \log\left(\frac{r}{a}\right).
\end{split}
	\label{sm:f^0}
\end{equation}
Applying the boundary conditions, we recover Saffman's result
\begin{equation}
	\psi^{(0)}(r,\theta) = -\frac{Ur\sin\theta}{\mathcal{C}}\left[ \mathcal{C} 
	+ \frac{1}{2} - \frac{a^2}{2r^2} - \log\frac{r}{a}\right],
	\label{sm:psi^0_saff}
\end{equation}
by design.

\subsubsection*{First order}
At first order, $\psi^{(1)}$ also satisfies the biharmonic equation, and 
therefore $\phi^{(1)} \sim \mathcal{C}_1/r + \mathcal{C}_2 r + \mathcal{C}_3 
r^3 + \mathcal{C}_4 r \log\left(r/a\right)$, as above.  However, applying the 
boundary conditions gives $\phi^{(1)}=0$ and therefore $\psi^{(1)} = 0$.

\subsubsection*{Second order}
At second order, $\psi^{(2)}$ satisfies an {\it inhomogeneous} biharmonic 
equation, which, in terms of the radial dependence, gives
\begin{equation}
	\hat{B}^2 \phi^{(2)} = \Phi,
	\label{sm:inhom}
\end{equation}
where
\begin{equation}
	\Phi := -\left\{\hat{B},\Delta^{(2)}\right\}\phi^{(0)} - 2K^{(2)} 
	\hat{B}\phi^{(0)} - 2\left\langle\nabla K^{(2)},\nabla 
	\phi^{(0)}\right\rangle.  \label{sm:Phi}
\end{equation}
Here, comma separated curly brackets $\{\cdot,\cdot\}$ are used to represent 
the anti-commutator, {\it i.e.}, 
$\{\hat{a},\hat{b}\}:=\hat{a}\hat{b}+\hat{b}\hat{a}$, for two operators 
$\hat{a}$ and $\hat{b}$.  The solution to (\ref{sm:Phi}) is a sum of 
homogeneous and particular parts, $\phi^{(2)}_h$ and $\phi^{(2)}_p$, 
respectively.  Here, $\phi^{(2)}_h$ is just the solution of the homogeneous 
equation, whilst $\phi^{(2)}_p$ can be found by constructing a Green's function 
$G(r,\xi)$ in order to satisfy the following equation:
\begin{equation}
	\hat{B}^2 \left[G\left(r,\xi\right)\right] = \delta\left(r-\xi\right).
	\label{sm:Greens}
\end{equation}
The right-hand side is just a Dirac delta function, and the operator $\hat{B}$ 
is given by the definition (\ref{sm:B_k}). For all values of $r$ other than 
$r=\xi$--- {\it i.e.}, intervals $a\leq r < \xi$ and $\xi < r \leq\infty$--- 
the function $G(r,\xi)$ satisfies the homogeneous equation $\hat{B}^2 
\left[G\left(r,\xi\right)\right] = 0$.  Therefore, applying the boundary 
conditions $G(a,\xi)=\mathrm{const}$ and $\lim_{r\to \infty} G(r,\xi) = 0$ 
implies that
\begin{equation}
	G(r,\xi) = \left\{
		\begin{array}{lr}
			\mathcal{D}_2 r + \mathcal{D}_3 r^3 + \mathcal{D}_4 r 
			\log\left(r/a\right), & a\leq r < \xi,\\
			\mathcal{D}_1/r, &\xi < r \leq\infty.
		\end{array}\right.
	\label{sm:G_split}
\end{equation}
where $\mathcal{D}_1$, $\mathcal{D}_2$, $\mathcal{D}_3$ and $\mathcal{D}_4$ are 
functions of $\xi$ determined by continuity conditions imposed on $G$ and its 
derivatives.  Since $\hat{B}^2$ is a fourth order operator, we impose that 
third order derivatives $d^3 G(r,\xi)/dr^3$ are discontinuous at $r=\xi$.  
Furthermore, $G(r,\xi)$ is assumed continuous in the variable $r$, and so too 
are its first and second derivatives $d G(r,\xi) / dr$ and $d^2 G(r,\xi)/dr^2$.  
In order to quantify the discontinuity in the third order derivative of $G$, we 
multiply (\ref{sm:Greens}) by $r$ and integrate over the interval 
$[\xi-\varepsilon, \xi+\varepsilon]$, giving
\begin{equation}
	\int_{\xi-\varepsilon}^{\xi+\varepsilon} 
	r\hat{B}^2\left[G\left(r,\xi\right)\right] dr = \xi.
	\label{sm:int_Greens}
\end{equation}
The integrand splits into two parts, one of which is an exact differential and 
therefore can be integrated easily, implying
\begin{equation}
	\begin{split}
	\xi =&\left[r\frac{d^3 G(r,\xi)}{dr^3} + \frac{d^2 G(r,\xi)}{dr^2} 
	-\frac{2}{r}\frac{d G(r,\xi)}{dr} + \frac{2}{r^2} 
G(r,\xi)\right]_{r=\xi-\varepsilon}^{r=\xi+\varepsilon}\\
&- \int_{\xi-\varepsilon}^{\xi+\varepsilon} 
\frac{1}{r}\hat{B}\left[G\left(r,\xi\right)\right] dr.
\end{split}
\label{sm:int_Greens_2}
\end{equation}
Taking the limit $\varepsilon\to 0$ of (\ref{sm:int_Greens_2}) then gives
\begin{equation}
	\lim_{\varepsilon\to 0}\left[\frac{d^3 
	G(r,\xi)}{dr^3}\right]_{r=\xi-\varepsilon}^{r=\xi+\varepsilon} = 1,
	\label{sm:G_discont}
\end{equation}
as required.  Using (\ref{sm:G_discont}) and the fact that
\begin{equation}
	\lim_{\varepsilon\to 0}\left[\frac{d^n 
	G(r,\xi)}{dr^n}\right]_{r=\xi-\varepsilon}^{r=\xi+\varepsilon} = 0,\ 
	\forall\ n=0,1,2,
	\label{sm:G_cont}
\end{equation}
Eq.~(\ref{sm:G_split}) can be solved for the functions $\mathcal{D}_1$, 
$\mathcal{D}_2$, $\mathcal{D}_3$ and $\mathcal{D}_4$.  The result is that
\begin{equation}
	\mathcal{D}_1=\frac{\xi^4}{16},\ 
	\mathcal{D}_2=\frac{\xi^2}{4}\log\left(\xi/a\right),\ 
	\mathcal{D}_3=\frac{1}{16},\ \mathrm{and}\ \mathcal{D}_4=-\frac{\xi^2}{4},
	\label{sm:D_n}
\end{equation}
which, when substituted into (\ref{sm:G_split}), gives the final form of our 
Green's function.  Convolving $G(r,\xi)$ with $\Phi$ then gives the result
\begin{equation}
	\begin{split}
	\phi^{(2)}_p =& \frac{1}{16 r}\int_a^r \xi^4 \Phi (\xi) \, d\xi
		+\frac{r}{4} \int_r^{\infty } \xi^2  \Phi (\xi ) \log (\xi/a ) \, 
		d\xi\\
		&+\frac{r^3}{16} \int_r^{\infty } \Phi (\xi ) \, d\xi -\frac{r}{4} \log 
		\left(r/a\right)\int_r^{\infty } \xi^2  \Phi (\xi ) \, d\xi,
	\end{split}
		\label{sm:phi_p}
\end{equation}
Finally, we remark that the function $\Phi$ is known, and can be written in 
terms of the functions $\phi^{(0)}$, $h$, and their derivatives.  Of course, 
$h$ is know (see main article) and $\phi^{(0)}$ is given by 
(\ref{sm:psi^0_saff}), however, the resulting form, in terms of Bessel 
functions of the second kind, is both cumbersome and not very illuminating.  It 
is included only for completeness.
%
%
\begin{widetext}
\begin{equation}
	\begin{split}
	\Phi&=
	\frac
		{V}
		{16\,\mathcal{C}\, r^4 \left[K_1\left(a/l\right)\right]^2}
	\Bigg\{
		-3\,\sigma\,r^2\,\left[ K_0\left(r/l\right) \right]^2
		\bigg[
			11\,a^2 + \left( 5+6\,\mathcal{C} \right)r^2 - 
			6\,r^2\,\log\left(r/a\right)
		\bigg]\\
	&\quad
	+
	\sigma\,r^2 \left[ K_2\left(r/l\right) \right]^2
	\bigg[ -7\,a^2 + \left( 2\,\mathcal{C} -9 \right)r^2 
	-2\,r^2\,\log\left(r/a\right) \bigg]\\
	&\quad
	-\frac{4\,r}{l} K_0\left(r/l\right)\,K_1\left(r/l\right)
	\bigg[ 3\left( 5 + 6\,\mathcal{C} \right)\kappa\, r^2 + 4\left( 
		2\,\mathcal{C} -1\right)\sigma\,r^4
	+ \left( 41\,\kappa + 4\,\sigma\,r^2 \right) a^2 - \left(9\,\kappa + 
	4\,\sigma\,r^2 \right)2r^2\,\log\left(r/a\right)
\bigg]\\
	&\quad
	-
	4 \left[ K_1\left(r/l\right) \right]^2 \bigg[ 3\left(5 + 
		6\,\mathcal{C}\right)r^2\,\kappa + 4\left( 1+2\,\mathcal{C} \right) 
		\sigma\,r^4 + \left( 49\,\kappa+12\,\sigma\,r^2 \right)a^2
	- \left( 9\,\kappa + 4\,\sigma\,r^2 \right)2\,r^2\,\log\left(r/a \right)
\bigg]
\Bigg\}.
\end{split}
	\label{sm:big}
\end{equation}
\end{widetext}
Turning attention to the homogeneous part, a straightforward application of the 
boundary conditions gives $\phi^{(2)}_h = \mathcal{A}/r + \mathcal{B}$, 
%
%
where $\mathcal{A}$ and $\mathcal{B}$ are the following constants:
\begin{equation}
	\begin{split}
	\mathcal{A}=&\frac{a^2}{2} + \frac{3 a^4}{16} \int_a^{\infty } \Phi (\xi ) 
	\, d\xi +\frac{a^2}{4} \int_a^{\infty } \xi^2  \Phi (\xi ) \log (\xi/a ) \, 
	d\xi\\
	&- \frac{a^2}{4} \int_a^{\infty } \xi^2  \Phi (\xi ) \, d\xi,
\end{split}
	\label{sm:Aconst}
\end{equation}
\begin{equation}
	\begin{split}
	\mathcal{B}=& -\frac{a^3}{4}\int_a^{\infty } \Phi(\xi ) \, d\xi 
	-\frac{a}{2} \int_a^{\infty } \xi^2  \Phi(\xi ) \log (\xi/a ) \, d\xi\\
	&+ \frac{a}{4}\int_a^{\infty } \xi^2  \Phi(\xi ) \, d\xi.
	\end{split}
	\label{sm:Bconst}
\end{equation}
Notice that, as required, $\lim_{r\to\infty}\phi^{(2)}(r) = 0$ and 
$\phi^{(2)}(a) = Va/2$ [{\it cf.} Eqs.~(\ref{sm:bound_psi^0^1}) and 
(\ref{sm:bound_psi^2})].  Due to the complexity of $\Phi$, the integrals 
involved in the above expressions [(\ref{sm:phi_p}), (\ref{sm:Aconst}) and 
(\ref{sm:Bconst})] cannot be solved analytically.  Therefore, at second order, 
we are restricted to numerical solutions.  (Albeit ones that are relatively 
straightforward to compute).

\subsection*{Lateral mobility}
Since, in general, we are restricted to perturbative solutions for $\psi$, the 
equation for the drag coefficient must also be considered order by order:
\begin{equation}
	F^{(0)} + \alpha F^{(1)} + \alpha^2 F^{(2)} = -\left[\lambda^{(0)} + \alpha 
	\lambda^{(1)} + \alpha^2 \lambda^{(2)} \right] V + O\left(\alpha^3\right).
	\label{sm:drag_perturb}
\end{equation}
Here, the left hand side corresponds to Eq.~(\ref{sm:drag_2}) which, by 
substituting for the definition of the stress tensor $\Pi^{ij}$, can be written 
as
\begin{equation}
	F = \int_{\partial\mathcal{S}} 
	\left(\hat{\boldsymbol{i}}\cdot\boldsymbol{e}_i\right)\left[ - p g^{ij} + 
	\eta\left( v^{i;j} + v^{j;i}\right) \right]\,dl_j.
	\label{sm:drag_3}
\end{equation}
In principle, we now have everything necessary to expand this expression in 
terms of $\alpha$ and compute the coefficients.  That is, from the definition 
of the tangent basis $\boldsymbol{e}_i$ [see Eq.~(\ref{sm:tangent_vecs})], it 
is clear that
\begin{equation}
	\hat{\boldsymbol{i}}\cdot\boldsymbol{e}_1 = \cos\theta\ \mathrm{and}\ 
	\hat{\boldsymbol{i}}\cdot\boldsymbol{e}_2=-r\,\sin\theta.
	\label{eqidotei}
\end{equation}
Furthermore, from (\ref{sm:line}), the line 1-form(s) $dl_i$ are given by 
\begin{equation}
	dl_1 = \left[ 1 + \alpha^2 \frac{\left( h' \right)^2}{2}  + O\left( 
	\alpha^3 \right)\right] r\,\mathrm{d}\theta,\ \mathrm{and}\ dl_2 = 0.
	\label{sm:line_i}
\end{equation}
All that remains, therefore, is to calculate the components of the stress 
tensor from the results in the previous section.  The coefficients, in a series 
expansion in terms of $\alpha$, of the component functions $v^i$ are given by 
Eqs.~(\ref{sm:v_stream_0_1}) and (\ref{sm:v_stream_2}).  Calculating the 
covariant derivative [see Eq.~(\ref{sm:cov_components})] of these functions 
then requires the Christoffel symbols (\ref{sm:Gamma_pert_1}) and 
(\ref{sm:Gamma_pert_2}).  Similarly, the coefficients in the $\alpha$-expansion 
of the pressure
\begin{equation}
	p=p^{(0)} + \alpha\,p^{(1)} + \alpha^2\,p^{(2)} + O\left(\alpha^3\right).
	\label{sm:p_pert}
\end{equation}
may be calculated by combining (\ref{sm:post_permute}) with the series 
expansions (\ref{sm:psi_pert}) to (\ref{sm:K_pert}).

\subsubsection*{Zeroth order}
In a straightforward application of Eqs.~(\ref{sm:v_stream_0_1}) and 
(\ref{sm:v_stream_2}), the lowest order coefficients of the velocity functions 
can be easily calculated.  As can the lowest order coefficient of the series 
expansion of the pressure.  In both cases, we simply reproduce the results of 
Saffman \cite{PGS75} and the resulting drag is therefore
\begin{equation}
	\lambda^{(0)} = \frac{4\pi\eta}{\mathcal{C}}.
	\label{sm:lambda^0}
\end{equation}

\subsubsection*{First order}
By virtue of the fact that $\psi^{(1)}=0$, the definitions 
(\ref{sm:v_stream_0_1}) and (\ref{sm:v_stream_2}) imply that, at first order, 
the coefficients (in the $\alpha$-expansion) of both the velocity components 
and the pressure are zero {\it i.e.},
\begin{equation}
	\left[ v^i \right]^{(1)}=0,\ \forall\ i=1,2,\ \mathrm{and}\ p^{(1)}=0.
\end{equation}
Notice that this does not mean that there is no first order correction to the 
velocity $\boldsymbol{v}=v^i\,\boldsymbol{e}_i$, because of the 
$\alpha$-dependence of the basis vector $\boldsymbol{e}_1$ [{\it cf.} 
Eqs.~(\ref{sm:tangent_vecs})].  However, this correction plays no role in the 
net force $F$, which acts opposite to the direction of movement (i.e., 
in-plane).  That is
\begin{equation}
	F^{(1)} = 0,\ \Longrightarrow\ \lambda^{(1)}=0.
	\label{sm:lambda^1}
\end{equation}
This should not be a surprise as the membrane is up/down symmetric and 
therefore any corrections to the drag coefficient $\lambda$ should be invariant 
under the sign change $\alpha\rightarrow-\alpha$.

\subsubsection*{Second order}
As explained above, in principle, the solution to $\phi^{(2)}$, combined with 
the trivial angular dependence of $\psi^{(2)}$, is enough to calculate the 
components of the velocity field [via (\ref{sm:v^i_pert})] and then the 
pressure field [via (\ref{sm:post_permute})].  In practice, the manipulations 
are extremely tedious.  We therefore used the commercially available symbolic 
computation software, Mathematica \cite{Wolfram}.  The form of the resulting 
expressions are not particularly illuminating.  Nevertheless, for completeness, 
we state the final expression for the drag coefficient at second order below.  
(The reader is reminded that, due to the complexity of the function $\Phi$, 
this expression must be calculated numerically).
\begin{widetext}
\begin{equation}
	\begin{split}
		\lambda^{(2)}=\frac
	{
		\eta\,\pi
	}
	{
		4\,\mathcal{C}\,l^2
		\left[
			K_1\left( a / l \right)
		\right]^2
	}
	\Bigg\{&
		3\,a^2\,\mathcal{C}
		\left[
			K_0\left( a/l \right)
		\right]^2
		+
		a^2\,\mathcal{C}
		\left[
			K_2\left( a/l \right)
		\right]^2
		+
		12\,a\,l
		\left(
			\mathcal{C} + 2
		\right)
		K_0\left( a/l \right)
		K_1\left( a/l \right)\\
		&+
		\left[
			K_1\left( a/l \right)
		\right]^2
		\Bigg[
			4
			\left(
				a^2\,\mathcal{C}
				+ 3\,l^2\,(4+\mathcal{C})
			\right)
			+
			\frac{\mathcal{C}\,l^2}{V}
			\int_a^\infty \left[7\xi^2 - 3\,a^2 - 
			\xi^2\,\log\left(\xi/a\right)\right]\,\Phi(\xi)\,d\xi
		\Bigg]
	\Bigg\}.
	\end{split}
	\label{sm:lambda^2}
\end{equation}
\end{widetext}

\end{document}